\documentclass[12pt,preprint]{aastex}

\shorttitle{Outflow Driven Turbulence}
\shortauthors{Carroll, Frank, Blackman, Cunningham, Quillen}

\begin{document}
\title{Outflow Driven Turbulence in Molecular Clouds}

\author{Jonathan J. Carroll\altaffilmark{1}, Adam Frank\altaffilmark{1}, Eric G. Blackman\altaffilmark{1}, Andrew J.
Cunningham\altaffilmark{1,2}, Alice C. Quillen\altaffilmark{1}}
\altaffiltext{1}{Department of Physics and Astronomy, University of Rochester, Rochester, NY 14620}
\altaffiltext{2}{Lawrence Livermore National Laboratory, Livermore, CA 94550}
\email{johannjc@pas.rochester.edu}

\begin{abstract}
In this paper we explore the relationship between protostellar outflows and turbulence in molecular clouds. Using 3-D
numerical simulations we focus on the hydrodynamics of multiple outflows interacting within a parsec scale volume.  We
explore the extent to which transient outflows injecting directed energy and momentum into a sub-volume of a molecular cloud can be converted into random turbulent motions.   We show that turbulence can readily be sustained by these
interactions and show that it is possible to broadly characterize an effective driving scale of the outflows.  We
compare the velocity spectrum obtained in our studies to that of isotropically forced hydrodynamic turbulence finding
that in outflow driven turbulence a power law of the form $E(k) \propto k^{-\beta}$ is indeed achieved. However we find a steeper spectrum $\beta \sim 3$ is obtained in outflow driven turbulence models than in isotropically forced simulations $\beta \sim 2.0$.  We discuss possible physical mechanisms responsible for these results as well and their implications for turbulence in molecular clouds where outflows will act in concert with other processes such as gravitational collapse.
\end{abstract}


\section{Introduction} \label{intro}
The majority of stars form in self gravitating cores within giant molecular clouds (GMCs).  Two important issues facing
modern theories of star formation are the nature of turbulence in these clouds and the relative inefficiency of star
formation.  For excellent reviews see \cite{ElmegreenScalo,McKeeOstriker}.  These two issues are likely to be
related.  The star formation rate per free fall ($SFR_{ff}$) can be defined as the fraction of mass that is converted
into stars within one free fall time at the mean density $SFR_{ff}=\frac{\dot{M}_{\star}t_{ff}}{M}$
\citep{krumholz-turb-reg}. In the absence of some form of support, most of the mass in GMC's would collapse into stars
within a free fall time $t_{ff} = [3\pi(32G\rho_0)]^{1/2}$ giving an $SFR_{ff}\sim{} 1$.  Observations however find
surprisingly low values of $SFR_{ff}$'s (typically ranging from .01 to .1).  Theoretical accounts for the low values of
$SFR_{ff}$ will rely on some form of support within the cloud to keep it from collapsing to form stars. Supersonic
turbulence is one means by which this can be achieved. Turbulence in molecular clouds is inferred from Larson's Laws which
are empirical relationships between line-widths and size observed in many star forming regions \citep{larson}.
Turbulence has also been inferred from direct measurements of power spectra in molecular clouds across a wide range
of scales \citep{heyer}.  While turbulence can provide an isotropic pressure to support the cloud against self-gravity,
both hydrodynamic and MHD turbulence decay quickly \citep{stone,maclow-dissipation}.   Thus if clouds are stable,
long-lived structures supported against self-gravitational collapse by turbulence, those turbulent motions must 
be continually driven either internally via gravitational collapse and stellar feedback or externally via turbulence in
the ISM.

Feedback in the form of stellar outflows is one means of driving turbulence in molecular clouds \citep{normansilk}.  If
outflows are well coupled to the cloud then observational studies make it clear that combined outflow energy budgets
are sufficient to account for cloud turbulent energy \citep{bally96,ballyreview,knee,quillen,warin}.  Analytical work
by \citet{matzner00,matzner01,matzner} has explored the role of collimated outflow feedback on clouds.
\cite{matzner}, in particular, developed a theory for outflow driven turbulence in which line-widths were predicted as
functions of a global outflow momentum injection rate.  \cite{krumholz} have also considered the nature of feedback via
outflows, concluding that these systems provide an important source of internal driving in dense star forming cores.
Early simulation studies also indicated that outflows could drive turbulent motions \citep{maclow-outflows}.  More
recent work by \cite{linaka} and \cite{nakamura} have mapped out the complex interplay between star formation and turbulence concluding that outflows could re-energize turbulence.  However, studies of single jets by
\cite{banerjee} came to the opposite conclusion in the sense that single jets will not leave enough supersonic material
in their wakes to act as a relevant source of internal forcing.

In this letter we address the issue of outflow driven turbulence by focusing solely on the dynamics of cloud material
set in motion by multiple transient collimated jets.  We use a higher resolution of the outflows than previous studies
to tease out the role of their dynamics in driving turbulence. Following the dimensional analysis of \cite{matzner} we
consider a cloud of mean density $\rho_0$, with outflows occurring at a rate per volume $\mathcal{S}$ with momentum
$\mathcal{I}$.  This then defines characteristic outflow scales of mass, length, and time:
\begin{equation}
\mathcal{M}=\frac{\rho_0^{4/7}\mathcal{I}^{3/7}}{\mathcal{S}^{3/7}},
\mathcal{L}=\frac{\mathcal{I}^{1/7}}{\rho_0^{1/7}\mathcal{S}^{1/7}},
\mathcal{T}=\frac{\rho_0^{3/7}}{\mathcal{I}^{3/7}\mathcal{S}^{4/7}}
\end{equation}
Combining these gives other characteristic quantities.  Of particular interest is the characteristic
velocity:
\begin{equation}
\mathcal{V}=\frac{\mathcal{L}}{\mathcal{T}}=\frac{\mathcal{I}^{4/7}\mathcal{S}^{3/7}}{\rho_0^{4/7}}=\frac{\mathcal{I}}{\rho_0
\mathcal{L}^3}
\end{equation}
Assuming typical values for $\rho_0$, $\mathcal{I}$, and $\mathcal{S}$ yields supersonic characteristic velocities
indicating that outflows contain enough momentum to drive significant supersonic turbulence.  Our simulations seek to
explore a detailed realization of this idea.  We note that these relations are for spherical outflows and
provide appropriate order of magnitude estimates for collimated outflows.

This work is the continuation of a series of studies on outflows and turbulence. In \cite{quillen} observations of
dense gas in NGC 1333 showed numerous fossil outflow cavities with total energy and momenta consistent with
re-energizing turbulence \citep{cunningham-cavity}. Most recently \cite{cunningham-sj} have carried out simulations of a single outflow cavity interacting with a turbulent medium.  They conclude that outflow driven cavities are able to
re-energize turbulent motions in their immediate environment provided such turbulent motions already exist to disrupt
the cavity.  Here we numerically investigate the degree to which interacting outflow cavities are able to provide such
a turbulent environment in regions typical of star forming clusters.  Future studies will include treatment of
additional physics including magnetic fields (Carroll in prep.), self-gravity and radiation transport.  We note the
studies are relevant to broader question of turbulence driven by continuous vs discrete forcing \citep{JoungMacLow}

In section~\ref{numerical_model} we describe the numerical model and the properties of individual outflows, in
section~\ref{results} we discuss the characteristics of outflow driven turbulence, and in section~\ref{conclusion} we
summarize the implications of outflow driven turbulence within molecular clouds.

\section{Numerical Model} \label{numerical_model}
In this study we use a new Reimann solving MHD code called AstroCUB.  The code is 2nd order accurate in space and time
and uses a non-split CTU method as described by \cite{GardinerStone}.  Our current simulations were purely hydrodynamic
and were performed on a periodic cube of length 1.5 pc for 1 Myr at a resolution of $256^3$ using a polytropic equation
of state ($\gamma=1.0001$) to approximate an isothermal gas at 10 \degr K.  Highly collimated bipolar outflows of
momentum $\mathcal{I}=10^{39.5}$ g cm s$^{-1}$ were launched at a constant rate $\mathcal{S}=10^{-67.2}$ cm$^{-3}$
s$^{-1}$ into an initially uniform environment of density $\rho_0=10^{-19.6}$ g cm$^{-3}$.  The values for
$\mathcal{I}$, $\mathcal{S}$, and $\rho_0$ were chosen to approximate star forming regions like NGC1333.  Using the
scaling relations above, these define outflow scales of length $\mathcal{L}=.36 pc$, mass $\mathcal{M}=17.34 M\sun$,
and time $\mathcal{T}=.366 Myr$.  The simulation domain was chosen to be 4 times the outflow length scale so the
corresponding outflow wave number $\mathcal{K}=\frac{2\pi}{\mathcal{L}} = 4\times k_{min}$ where $k_{min}$ corresponds
to the scale of the box.  Outflows were continually launched for 3 times the outflow time scale for a total of $4^3
\times 3 = 192$ outflows launched during the simulation.  The turbulence was then allowed to decay for another 3.7 Myr
which corresponds to 2.4 dynamical times ($t_d=\frac{L_{box}}{\mathcal{V}}$).  In addition to the outflow-driven
turbulence case we also carried forward a ``standard'' isotropically forced turbulence simulation as a control case. In
this simulation turbulent motions where imposed at each timestep using a series of sinusoidal driving functions with
an ``injection'' or ``driving'' scale of $\mathcal{K}= 4\times k_{min}$

Each outflow was launched with a $5\degr$ half opening angle from a region 10 cells across corresponding to an outflow
radius of 5800 AU.  The velocity, density, and duration of each jet were modeled on a $.5 M_{\sun}$ star ejecting
one-sixth of its mass at a velocity of 240 km/s giving $v_c = \frac{\mathcal{I}}{M_{\star}}$=40 km/s.  The mass loss
rate was chosen to be $10^{-4}$M$_{\sun}$/yr giving a jet density of .21$\rho_0$ and duration of .83 kyr.  The actual
outflows injected less momentum (80\%) and about twice the mass, however the total mass injected by all 192 outflows
amounts to less than 3\% of the ambient material and does not have a significant effect on the overall dynamics.

\begin{figure}[hb!]
\begin{center}
  \includegraphics[width=.9\textwidth]{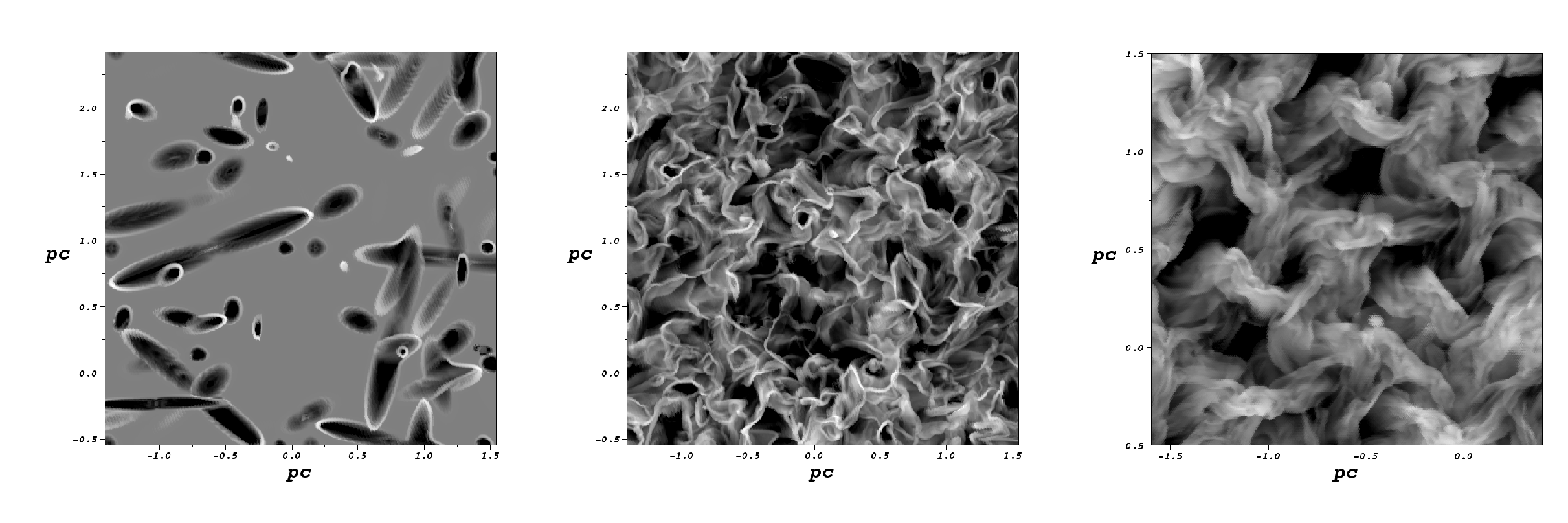}
\end{center}
\caption{Cross cuts of log density for the outflow driven turbulence at $.5\mathcal{T}$ (Left), and at $2.6\mathcal{T}$
(Center) as well as for the isotropically forced turbulence (Right).  Note the interaction of outflow shells at early
times (Left) and the presence of distinct ``holes'' in the turbulent density distribution at later times (Center).  (The colorbars for the outflow driven turbulence and the isotropically forced turbulence cover the ranges [.05-20]$\rho_0$ and [.1-10]$\rho_0$ respectively)}
\label{logdcc}
\end{figure}

\section{Results} \label{results}
\subsection{Creation and Maintenance of Turbulence}
Figure \ref{logdcc} shows density cross cuts from both the outflow driven and isotropically driven control
simulations.  Examination of the outflow driven simulation shows that initially outflow cavities are able to expand into the quiescent environment without being disrupted and easily grow to pc size structures.  Since the outflows are
bipolar, the total vector momentum injected by each outflow is zero, however, the total scalar momentum defined as
$\int{\left|\mathbf{P}\right|dV}$ grows steadily until the cavity interactions begin to dissipate momentum in the
expanding shells. By $1.8\mathcal{T}$ the cavities have entirely filled the domain and by $3\mathcal{T}$ the system has reached a statistically quasi-steady state.  Comparison of the late time outflow driven simulation image and the
isotropically forced simulation image show that both have reached states of highly disordered flows with structure
present on a variety of scales. As we show below the visual impression of turbulence in both cases is supported by
statistical measures of the flows as well as their time evolution.  The comparison by visual inspection however is noteworthy because of the ``holes'' which appear in the outflow driven density distribution. These are created by fossil outflow cavities and exist until the cavities are subsumed by the turbulent motions \cite{cunningham-sj} or interact with another cavity.  Such shells have been observed in turbulent flows around star forming regions such as NGC1333
\citep{quillen} and point to an important morphological, rather than statistical, signature of outflow driven
turbulence.

\begin{figure}[ht!]
\begin{center}
\begin{tabular}{l}
  \includegraphics[width=.4\textwidth]{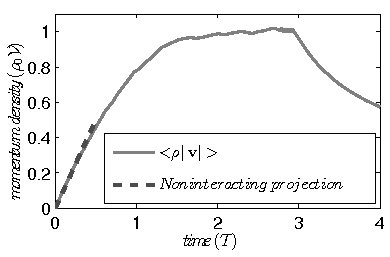}
  \includegraphics[width=.4\textwidth]{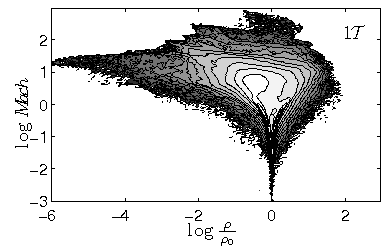} \\
  \includegraphics[width=.4\textwidth]{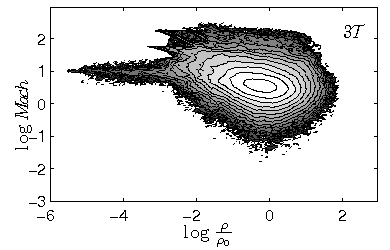}
  \includegraphics[width=.4\textwidth]{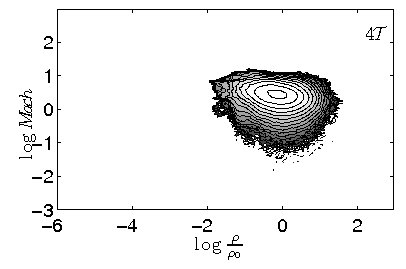} \\
\end{tabular}
\end{center}
\caption{Shown here are the time evolution of the total scalar momentum (upper left) and two-dimensional pdf's of the
gas density and Mach number at t=1,3,\&4 $\mathcal{T}$ showing the saturation and decay of the supersonic turbulence.  The pdf's are filled contours equally spaced in log density with the fill color corresponding to the probability density (white being the maximum).
Note that no jets are launched after 3$\mathcal{T}$.}
\label{2dpdfs}
\end{figure}

In figure \ref{2dpdfs} we show a plot (upper left panel) of the average scalar momentum density vs time.  The scalar
momentum density and time have been scaled to $\rho_0\mathcal{V}$ and $\mathcal{T}$ respectively.  Initially the scalar momentum density injection rate is $\mathcal{I}\mathcal{S}=\frac{\rho_0\mathcal{V}}{\mathcal{T}}$ since the jets are
not interacting.  Around $.5\mathcal{T}$ the expanding cavities begin to interact and scalar momentum begins to be
dissipated.  By $2\mathcal{T}$ the momentum dissipation rate balances the momentum injection rate and the system
reaches a steady state. The average mass weighted velocity in our simulation is very close to $\mathcal{V}$
corresponding to mach 4.8 or $v \sim .96 km/s.$ At $3\mathcal{T}$ the jets are turned off, and the scalar momentum
density dissipation rate is shown to be approximately $\frac{\rho_0\mathcal{V}}{\mathcal{T}}$.

Figure \ref{2dpdfs} also show the time development of the two-dimensional probability density functions for the gas
density and mach number.  The jets have stirred up all of the ambient material by $\sim 2\mathcal{T}$ and have reached
a statistically steady state by $\sim 3\mathcal{T}$.  The steady state density and velocity pdf's are fairly log normal
apart from the spikes associated with low density, high velocity material from successive outflow cavities.  Once the
jets are turned off at $3\mathcal{T}$ the turbulence decays fairly rapidly.  Note that by $4\mathcal{T}$ the pdf's show
no evidence of the expanding outflow cavities and are characteristic of a purely turbulent isotropic media.  The
development of log-normal density distributions in isothermal conditions is one measure of a turbulent flow
\citep{Vazquez-Semadeni(1994)}. The four panels in Figure~\ref{2dpdfs} demonstrate one of the key conclusions of our study: {\it transient outflow cavities can set the bulk of initially quiescent material into random but statistically steady supersonic motions.}

\subsection{Outflow driven turbulent spectra}
Supersonic turbulence in general involves a cascade of kinetic energy (or scalar momentum) from some forcing scale to smaller scales where energy is eventually dissipated in the form of heat and radiated away.  In between the forcing scale and the dissipation scale lies the inertial range where the turbulence is scale free and the energy spectra follows a power law $\mathcal{E}\left(k\right)\propto k^{-\beta}$.  Recent simulations of isothermal supersonic turbulence by \cite{kritsuk} have found that $\beta\sim2$ in the inertial range, and that the lack of eddies at the numerical dissipation scale ($\Delta x$) causes the spectrum to steepen for scales $\leq 8 \Delta x$.  This lack of eddies inhibits the cascade of energy from slightly larger scales creating a bottleneck that causes the spectra to flatten above $8 \Delta x$ up to approximately $32 \Delta x$.  Above $32 \Delta x$ the cascade is effectively inertial.  Modeling the cascade of energy in realistic
astrophysical conditions is therefore a difficult task as the numerical dissipation scale is often several orders of
magnitude above the physical dissipation scale in high Reynolds number flows characteristic of astrophysical
turbulence.

\begin{figure}[ht!]
\begin{center}
\begin{tabular}{l}
  \includegraphics[width=.4\textwidth]{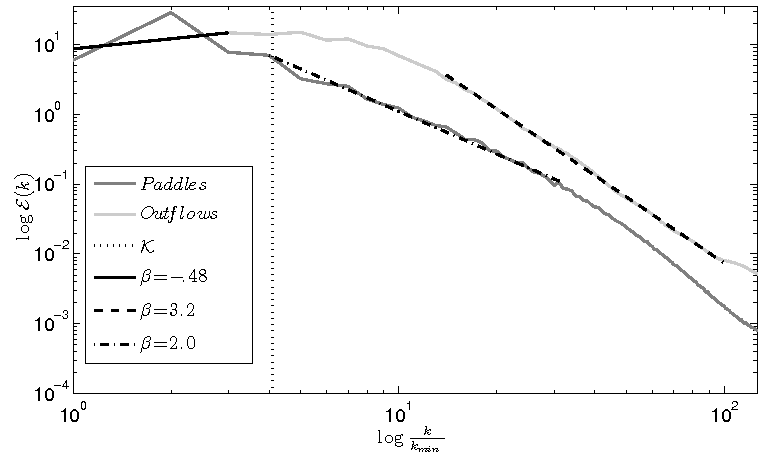}
  \includegraphics[width=.4\textwidth]{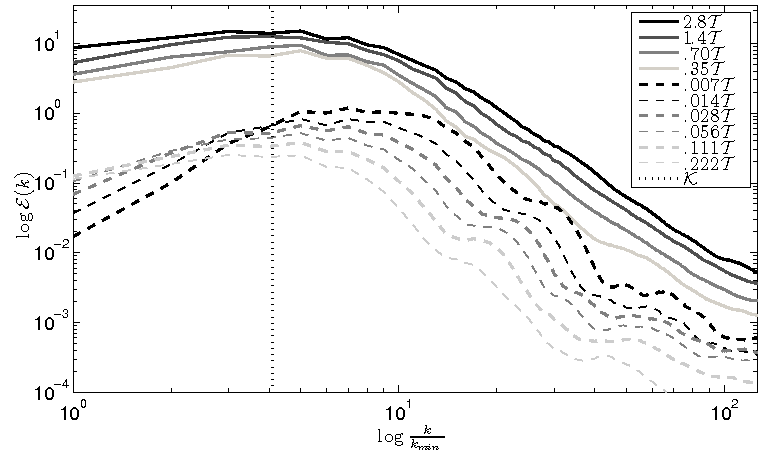}
\end{tabular}
\end{center}
\caption{The left panel shows a comparison of the velocity power spectrum for the outflow driven turbulence at $3
\mathcal{T}$ and isotropically forced turbulence with injection scale $[2.1 - 2.4] \mathcal{L}$.  The right panel shows
the development of the power spectra for both single and multiple outflow runs.  The characteristic wave number for the
outflows $\mathcal{K}=\frac{2\pi}{\mathcal{L}}$ is shown for reference.}
\label{spectra}
\end{figure}

Here we define the one-dimensional power spectra $\mathcal{E}$ of the velocity $\mathbf{u}$
\begin{eqnarray}
\mathcal{E}\left(k\right)\equiv\frac{1}{V}\int{\left|\tilde{\mathbf{u}}\left(\mathbf{k}\right)\right|^2\delta\left(\left|\mathbf{k}\right|-k\right)d\mathbf{k}}S
\\
\mbox{ where }\tilde{\mathbf{u}}\left(\mathbf{k}\right)=\int_V{\mathbf{u}\left(\mathbf{x}\right)e^{-2\pi
i\mathbf{k}\cdot\mathbf{x}}d\mathbf{x}} \\
\mbox{ so that }
\int{\mathcal{E}\left(k\right)dk}=\frac{1}{V}\int_V{\mathbf{u\left(x\right)}^2d\mathbf{x}}=\left<\mathbf{u}^2\right>
\end{eqnarray}
In figure \ref{spectra} we plot the one-dimensional velocity power spectra of the outflow driven turbulence as well as
the isotropically forced turbulence with injection scales $[2.1 - 2.4] \mathcal{L}$.  The spectrum of the isotropically forced turbulence follows a power law down to scales $\sim 8\Delta x$ ($k/k_{min} = 32)$ before the numerical dissipation overwhelms the bottleneck it produces.  We note of course that resolution is a critical issue in interpreting turbulent power spectra and we will address these issues in an subsequent paper (Carroll {\it et al.} in preparation).  For the present purposes the results allow us to compare the discrete source, outflow driven turbulence to the continuous, isotropically forced turbulence at the same resolution and mach number even though our inertial range is only marginally resolved.

Consideration of the right hand panel of Figure~\ref{spectra} demonstrates that the spectrum of the outflow driven
turbulence is quite different from the isotropically forced turbulence.  At large scales, the outflow driven spectrum rises slowly with $\beta=-.48$ from the box scale to around $\mathcal{K}$ in a broad maximum before turning over and steepening to $\beta=3.2$ at $\sim3\mathcal{K}$.  The power law behavior continues all the way to the dissipation scale.  The isotropically forced spectrum shows a sharp peak at the driving scale and then falls with shallower power law appropriate to a Burgers model for compressible turbulence in which the random distribution of shocks gives a spectrum with $\beta = 2$ without regard to the cascade of energy as in Kolmogorov turbulence.  \citet{matzner} suggested that a cascade of momentum in supersonic turbulence would also produce a $\beta=2$.

The reason for the difference between the outflow driven turbulent spectra and the spectra expected to be produced by shock-filled turbulence as well as a momentum cascade model, lies in the nature of the driving.  Since the shocks in the outflow driven turbulence are not randomly distributed (i.e. opposing bow shocks associated with bipolar outflows), the turbulence though shock-dominated, should not be expected to follow a Burgers model power law.  In addition, because the turbulent turnover time scale at the outflow scale $\frac{\mathcal{L}}{\mathcal{V}}=\mathcal{T}$ is equivalent to the time scale for an outflow to clear out a region of volume $\mathcal{L}^3$, the spectrum at scales below the outflow scale $k > \mathcal{K}$ would not be expected to follow a momentum type cascade model either.  The steep spectrum at sub outflow scales (with $\beta \sim 3$ corresponding to a velocity length scaling $v(l)\propto l$) is consistent with the Hubble type flow seen in the expanding cavities.  The time evolution of the spectra (for many outflows) shown in the right hand panel of figure~\ref{spectra} provides some insight. The steep slope of the spectrum at sub-outflow scales is present already by $.70\mathcal{T}$ before the outflows have begun to significantly interact.  This is consistent with the interpretation that the steep slope of the spectrum at later times $3\mathcal{T}$ arises from the volume swept-up by the multiple fossil outflow cavities rather than their interaction.  In other words, we can interpret the steep spectra as due to expanding outflows that sweep up small-scale (high $k$) eddies which effectively removes them from the flow.  This effect would reduce the observed power on these scales and steepen the slope of the energy spectrum.

\subsubsection{Large scale spectra and forcing functions}
In order to clarify the behavior of outflow driven turbulence, a single outflow simulation was run. In the right hand panel of figure~\ref{spectra} we present the development of the velocity power spectra for this single jet.  The six snapshots are equally spaced in log time.  For each line the velocity power spectra peaks at approximately the outflow length scale which grows in time. Note the slight secondary plateau at the outflow width scale.  As the expanding cavity grows, the total kinetic energy decreases, and the energy shifts to larger scales.  For an expanding
outflow in a quiescent medium there is, of course, nothing to limit the growth.  Outflows in a turbulent medium
however, will loose coherence when the momentum in the outflow is comparable to the turbulent momentum swept up by the
expanding cavity. This statement is equivalent to the fact that once the outflows reach the scale $\mathcal{L}$ they
interact and are destroyed, converting their directed motion up to random motion. For spherical outflows, this will happen on scales $\sim \mathcal{L}$.  Collimated outflows would in theory be able to grow to larger size structures before loosing coherence, however the momentum in the expanding sides of the cavities can be comparable to that in the head of the jet.  In fact we find that after 1, 2, \& 3 $\mathcal{T}$, the transverse momentum accounts for 30\%, 40\%, \& 50\% of the total momentum respectively.  So a driving scale $\geq \mathcal{L}$ is still a fairly good approximation.
One would expect the spectrum to fall off steeply for scales above the driving scale $k < \mathcal{K}$.  Both panels in figure~\ref{spectra} however, show that multiple outflows inject their energy over a broad range of scales, and while each outflow injects most of its energy at $L_{of} \sim \mathcal{L}(t)$, the shorter turnover times for scales closer to $\mathcal{L}$ cause the spectra to flatten making identification of a single outflow driving scale difficult.  To
address this issue we consider combining a power law forcing function $\mathcal{F} \propto k^\alpha$ with a power law energy
spectrum $\mathcal{E} \propto k^{-\beta}$ and a dissipation rate to smaller scales $\Pi \propto \frac{v_l^2}{t_l}
\propto kv_k^3 \propto k^{5/2}\mathcal{E}_k^{3/2} \propto k^{3/2(5/3-\beta)}$.  With these relations we can look for
the steady state solution for the energy at each scale:

\begin{eqnarray}
  \frac{d\mathcal{E}_k}{dt}=\mathcal{F}_k-\frac{d}{dk}\Pi_k=0
\end{eqnarray}

Using the above assumptions this equation implies $\alpha = 3/2(1-\beta)$.  In other words, for $\alpha < 3/2$ a
forcing function with a positive slope will result in an energy spectrum that would have a negative slope in the forcing region. This is likely the case in our simulations for scales close to, but shorter than, our nominal driving
scale. Recall that at the largest scales we found the outflow turbulent spectra had a positive slope with $\beta=-.48$).  Here a relatively steep forcing function $\mathcal{F}\propto k^{2.2}$ could peak at $\sim \mathcal{L}$ but
reproduce the observed energy spectrum at $L > \mathcal{L}$.  Thus while our simulations do not begin with a specified
forcing function and simply allow the interactions of outflow cavities to self-consistently establish a turbulent flow
we can still draw broad conclusions about the spectral nature of the imposed forcing.

\section{Conclusion} \label{conclusion}
There have been many simulations of turbulence in both the atomic and molecular ISM
\citep{stone,kritsuk,porter,maclowkl}.  Some of these focus on decaying turbulence while others explore driven
turbulence.  In the majority of these studies the turbulence is forced with a continuous, isotropic and predetermined spectra.  In this paper we have focused on turbulence in sub-volumes of molecular clouds and considered the more realistic case of discrete, non-isotropic forcing in physical space rather than Fourier space.  Our goal has been to address the ability of transient protostellar outflows, ({\it fossil cavities}), to drive their surrounding media into supersonic turbulent motions.  Our results show conclusively that for reasonable values of outflow momenta $\mathcal{I}$ and outflow rate per unit volume $\mathcal{S}$, outflows generate supersonic turbulence.  We have found that the spectrum and physical properties of outflow driven supersonic turbulence differ from that produced by isotropic random forcing.  We note first that we are able to broadly identify an outflow forcing scale $\mathcal{L}$ which is of order of magnitude comparable with that predicted by \cite{matzner}. While the density pdf's are log-normal consistent with a isothermal turbulent flow \cite{Vazquez-Semadeni(1994)}, the spectra of outflow driven turbulence on scales smaller than $\mathcal{L}$ is steeper then that expected from turbulent cascade model as in isotropically forced turbulence.  We interpret this to be the result of overlapping outflow cavities preventing smaller scale structures from forming thereby facilitating a sort of bottleneck.  The small scale spectrum is therefore dominated by the spectrum of outflows cavities themselves which follow a power law consistent with a Hubble type flow. The departure from $\beta \sim 2$ velocity power spectra has been seen in other models with discrete forcing in the form of SN driven turbulence in the ISM \cite{JoungMacLow}. In those models a broad shallow plateau associated at large scales was found which then turned over to a steep spectrum at smaller scales which was attributed to numerical dissipation.

Our results add weight to the argument that protostellar outflows can play a significant role in altering their
star-forming environments \citep{nakamura}.  One must be careful not to over-interpret the current simulations which do
not include a number of important physical processes such as magnetic fields and self-gravity.  Our goal in this initial study was to focus on one key aspect of the physics at work in low mass star forming environments (multiple transient protostellar outflows) in order explicate its mechanisms and energetics.  The role of self-gravity, in particular, is likely to be important in determining the final turbulent state of the gas \citep{field} and magnetic fields may be able to transmit energy from regions of protostellar outflow to broader regions of the cloud \citep{KudohBasu}.  In spite of these important caveats, our results clearly establish a baseline for understanding how outflows can strongly effect stellar birth in clusters which are, after all, the regions where stars actually form.

\acknowledgments
We wish to thank Chris Matzner, Chris McKee and Mordecai-Mark Mac Low for extremely useful
discussions. Hector Arce, John Bally, Pat Hartigan and Tom Ray were also generous
with their time.  Tim Dennis, Kris Yirak, Brandon Schroyer and Mike Laski provided invaluable support and help.

Support for this work was in part provided by by NASA through awards
issued by JPL/Caltech through Spitzer program 20269, the National
Science Foundation through grants AST-0406823, AST-0507519 and
PHY-0552695 as well as the Space Telescope Science Institute through
grants HST-AR-10972, HST-AR-11250, HST-AR-11252.  We also thank the University of Rochester
Laboratory for Laser Energetics and funds received through the DOE
Cooperative Agreement No. DE-FC03-02NA00057.

\end{document}